\documentclass[useAMS,usenatbib]{mn2e}
\usepackage{epsfig}
\usepackage{amssymb}
\usepackage{amsmath}

\title[HV2112, a T\.ZO or a SAGB star]
{HV2112, a Thorne--\.Zytkow Object or a Super Asymptotic Giant Branch Star}

\author[C. A. Tout, A. N. \.Zytkow, R. P. Church \& H. H. B. Lau]
{Christopher A. Tout$^1$\thanks{email: cat@ast.cam.ac.uk}, Anna N. \.Zytkow$^1$,
Ross P. Church$^{2,1}$ and
\newauthor
Herbert H. B. Lau$^3$\\
$^1$Institute of Astronomy, The Observatories, Madingley Road,
Cambridge CB3 0HA\\
$^2$Department of Astronomy and Theoretical Physics, Lund Observatory,
Box 43, SE--221 00, Lund, Sweden\\
$^3$Argelander Institute for Astronomy, University of Bonn, Auf dem Huegel 71, D-53121 Bonn, Germany}

\begin{document}

\date{Accepted.  Received ; in original form} 
\pagerange{\pageref{firstpage}--\pageref{lastpage}} \pubyear{}

\maketitle

\label{firstpage}

\begin{abstract}

The very bright red star HV2112 in the Small Magellanic Cloud could be
a massive Thorne--\.Zytkow Object, a supergiant-like star with a
degenerate neutron core.  With its luminosity of over $10^5\,\rm
L_\odot$, it could also be a super asymptotic giant branch star, a
star with an oxygen/neon core supported by electron degeneracy and
undergoing thermal pulses with third dredge up.  Both T\.ZOs and SAGB
stars are expected to be rare.  Abundances of heavy elements in
HV2112's atmosphere, as observed to date, do not allow us to
distinguish between the two possibilities based on the latest models.
Molybdenum and rubidium can be enhanced by both the $irp$-process in a
T\.ZO or by the $s$-process in SAGB stars.  Lithium can be generated
by hot bottom burning at the base of the convective envelope in
either.  HV2112's enhanced calcium could thus be the key determinant.
A SAGB star is not able to synthesise its own calcium but it may be
possible to produce this in the final stages of the process that forms a
T\.ZO, when the degenerate electron core of a giant star is tidally
disrupted by a neutron star.  Hence our calculations indicate
that HV2112 is most likely a genuine T\.ZO.

\end{abstract}

\begin{keywords}
stars: AGB and post-AGB -- stars: abundances -- binaries: close --
stars: evolution -- stars: individual: HV2112
\end{keywords}

\section{Introduction}

Observations by \citet{levesque2014} of the very bright red star
HV2112 in the Small Magellanic Cloud (SMC) suggest that it could be a
Thorne--\.Zytkow object (T\.ZO).  \citet{thorne1975,thorne1977} first
modelled these stars with neutron-star cores and a structure somewhat
resembling that of a red supergiant.  \citet{levesque2014} demonstrate
that HV2112 is spectrally very different from other red supergiants in
the SMC.  It has noticeably enhanced spectral features corresponding
to rubidium and molybdenum, which were predicted to show up in massive
T\.ZOs \citep{Biehle1991,Biehle1994,cannon1993} as well as lithium
\citep{Podsiadlowski1995}.  It is not certain how these objects are
formed.  \citet{thorne1975,thorne1977} originally proposed either a
failed supernova or a neutron star accreting in a binary system.  In
the former case there is insufficient energy released to expel a very
massive envelope when a degenerate core collapses to a neutron star.
This is expected in the deaths of very massive stars
\citep{eldridge2004} but it is generally thought that fallback on to
the neutron star converts it to a black hole.  In the binary mechanism
the neutron star is formed in a normal supernova of the originally
more massive component in a close binary system.  The mass loss and
any kick are insufficient to unbind the system.  Subsequently its
companion evolves to a giant and fills its Roche lobe.  It is then the
more massive of the two objects and has a
deep convective envelope.  Mass transfer proceeds rapidly on a
time-scale approaching dynamical.  The core of the giant and the
neutron star are together both smothered by the giant's envelope and
spiral inwards to merge.  The last phase of the merge occurs once the
cores have spiralled close enough together that the electron
degenerate core is tidally disrupted and forms an accretion disc in
the orbital plane of the binary deep inside the envelope.  This occurs
on a very short time-scale and leaves a neutron star at the centre of
the giant common envelope.  This configuration settles to become the
T\.ZO.

Though the spectrum observed by \citet{levesque2014} differs
significantly from the other red supergiants they looked at in the SMC
they do not explicitly discuss super asymptotic giant branch (SAGB)
stars.  These are the late stages of stars of initial mass in a range
of a few solar masses somewhere between about 6~and $12\,\rm M_\odot$,
depending upon assumptions made about convective overshooting during
core helium burning, that have gone on to ignite carbon in their cores
before the second dredge up
\citep{GarciaBerro1994}.  In general, stars of
intermediate mass evolve through core hydrogen burning on the main
sequence.  When central hydrogen is exhausted, hydrogen burning
moves to a shell and the star becomes a red giant.  Its convective
envelope deepens and dredges some of the products of hydrogen burning
to the surface.  Once hot enough, helium burning led by the
triple-$\alpha$ process ignites in the core which burns convectively
to carbon and oxygen.  After its exhaustion in the core helium burns out
in a shell following the hydrogen burning shell.  These double shell
burning stars are on the asymptotic giant branch (AGB).  In the more
massive AGB stars, a second dredge up takes place when the deep
convective envelope penetrates beyond the temporarily extinct hydrogen
burning shell.  This brings new hydrogen fuel to reignite the hydrogen
shell only a few hundredths of a
solar mass outside the helium burning shell.  The thin extent of the
helium-rich region, coupled with the strong temperature sensitivity of
the triple-$\alpha$ reaction, causes unstable helium burning in pulses
between which episodes of third dredge up bring the products of helium
burning to the surface.  Amongst these are slow neutron capture
isotopes \citep{karakas2014} that can account for the heavier than
iron elements observed in HV2112.  The higher mass SAGB stars ignite
carbon before the second dredge up but go on to thermally pulse and
dredge up in a similar way.  They have higher core masses at second
dredge up and it is this that gives them luminosities as high as red
supergiants and T\.ZOs early on.  Once thermal pulsing and third dredge up has
begun, the core, and hence luminosity, grow much more slowly.

In the subsequent sections we look carefully at the various properties of
HV2112.  In almost all cases these are explained straightforwardly by both SAGB
stars and T\.ZOs.  The major exception is the enhanced calcium.  Neither SAGB
stars nor T\.ZOs can make calcium.  Its production by very hot 
burning may be due to more extreme processes involved in the formation of a T\.ZO.

\section {Luminosity and Temperature}

The bolometric luminosity of HV2112 is estimated to be $10^{5.15}\,\rm
L_\odot$.  The structure of a red giant envelope depends on the
luminosity generated deep with its core and its opacity which
determines its Hayashi track \citep{hayashi1961}.  Both T\.ZOs
\citep{cannon1993} and SAGB stars \citep{smartt2002,doherty2014} can reach
the required luminosity.  Because their structures, dense degenerate
cores and deep convective envelopes, are rather similar they appear at
similar locations in the Hertzsprung--Russell diagram.  In both cases
their luminosity is generated primarily by nuclear burning around their
compact degenerate cores.  Because both sorts of star lie on a Hayashi
track their temperature is almost entirely determined by their
luminosity with only a slight dependence on their total mass.  Normal
AGB stars can also reach these luminosities towards the end of their
lives but by then they are rich in heavy $s$-process elements such as
barium, and have destroyed all their lithium, inconsistent with HV2112
as we discuss below.

\section{Probabilities}

Neither SAGB stars nor T\.ZOs are common.  We make an order of magnitude
estimate of the number that should be expected in the SMC.  \citet{doherty2010}
find that, at SMC metallicity, SAGB stars form at masses between 6.5 and $8\,\rm
M_\odot$.  By a simple integration of the \citet{ktg93} mass function this
implies that one SAGB progenitor should form per $453\,\rm M_\odot$ of total
star formation.  The SAGB stage lasts for a few $10^5\,$yr while the total
lifetimes of the progenitors are a few $10^8\,$yr
\citep{doherty2010,doherty2014}.  Hence roughly one out of every thousand stars
of the correct mass should be a SAGB star at any given time.  From fig.~5 of
\citet*{glatt2010} we obtain that there are roughly 250~stellar clusters with
ages around 300\,Myr, and their fig.~13 suggests a mean mass of perhaps
$4000\,\rm M_\odot$, giving a total mass of about $10^6\,\rm M_\odot$ in
clusters that could host SAGB stars.  Putting these numbers together suggests
that the total number of SAGB stars in the SMC at the current time should be of
order unity.  Within the accuracy of this calculation this would be consistent
with HV2112 being the only SAGB star found in the SMC so far.

To compare the expected number of SAGBs with T\.ZOs we have used the binary star
population synthesis algorithm BSE \citep{hurley2002} to synthesise a population
of binary stars.  We give all stars above $8\,\rm M_\odot$ companions, with
masses chosen to be uniformly distributed in mass ratio $q$, and distribute the
initial semi-major axes uniformly in $\log a$ with $a$ between 10 and $10^4\,\rm
R_\odot$.  These are optimistic assumptions because we neglect non-interacting
wider binaries and any single massive stars.  We find that two per cent of our
binaries form a T\.ZO, equivalent to one T\.ZO per $10^{4}\,\rm M_\odot$ of
total star formation.  The progenitor lifetimes prior to T\.ZO formation are
typically of order 10\,Myr.  If we assume that the lifetimes of the T\.ZOs are
limited by strong winds driven by Mira-like pulsations with mass-loss rates
similar to the superwinds of AGB stars \citep{vassiliadis1993} then the
lifetimes of the T\.ZOs once formed are about $10^4\,\rm yr$.  So again roughly
one out of every thousand should be visible.  Taking there to be very roughly
$10^6\,\rm M_\odot$ in clusters of ages around $10^7\,{\rm yr}$ one would then
expect a probability of there being a T\.ZO visible in the SMC at the current
time of about ten per cent.  This number is in approximate agreement with the
calculation of \citet{Podsiadlowski1995} given the mass ratio of the SMC to the
Milky Way.  However, the rate of T\.ZO formation depends very strongly on the
assumptions made about the efficiency with which the envelope of the giant is
removed by the spiralling in process.  Here we assume that all the energy
liberated from the orbit goes into unbinding the envelope with no additional
energy, such as recombination of atoms, liberated.  Making different,
reasonable, assumptions we can easily change the formation rate by an order of
magnitude.  Hence it is not possible to draw any strong conclusions about the
likelihood of HV2112 being a T\.ZO from predicted formation rates alone.

It is also possible that T\.ZOs could be formed in the core-collapse
of the most massive single stars, above say $24\,\rm M_\odot$, if the
energy released is insufficient to eject all of the envelope.  In this
case there is no supernova explosion.  A simple calculation, analogous
to that for SAGB stars above, suggests that the resulting objects
should be roughly equally as common as SAGB stars.  If a sufficient
fraction of the material that has reached very high temperatures
during the core collapse can avoid being incorporated into the compact
object, this could potentially also produce the observed calcium
abundance.  It has always been assumed that, in this scenario,
material falling back on to the newly-formed neutron star would
convert it into a black hole.  This would destroy the star on a short
time-scale and prevent the $irp$-process from taking place.  However if
black hole formation can be avoided this mechanism would provide
roughly as many T\.ZOs as SAGB stars whilst allowing them to be rich
in calcium.

\section{Heavy Elements}

HV2112 is rich in rubidium and molybdenum.  These can be produced by
the $irp$-process in T\.ZOs \citep{cannon1993}.  They can also be
produced by the $s$-process in SAGB stars \citep[Doherty, private
  communication;][]{lau2011}.  The $s$-process begins with neutron
captures on to iron-group nuclei.  Initially lighter elements are
built up and both rubidium and molybdenum are members of this light
$s$-process set.  Subsequently, depending on the degree of neutron
exposure, heavier elements including barium and eventually lead are
formed.  \citet{levesque2014} found barium not to be enhanced in
HV2112 and claimed that this is evidence that the $s$-process is not
responsible for the rubidium and molybdenum.

The $s$-process nucleosynthesis of SAGB stars is still rather
uncertain.  \citet{lau2011} compute yields for 8, 8.5 and $9\,{\rm
M_\odot}$ stars of solar metallicity.  They find significant enhancements in
the light $s$-process elements consistent with neutrons produced by the
$^{22}\rm Ne(\alpha,n)^{25}Mg$ reaction.  The yields, defined as the ratios
of the average abundances in the stellar wind to those in the star initially,
are between a factor of $10^{0.5}$ and 10.  The enhancement in strong $s$-process
isotopes, however, depends very strongly on the rates of the triple-alpha and
$^{12}\rm C(\alpha,\gamma)^{16}O$ reactions, which control the temperature at
the base of the intershell convective zone.  Lower, older rates led to very
little heavy $s$-process, while newer, faster rates produce significant
quantities.  In light of this result and in the absence of a more detailed
analysis we consider that the abundance trends in rubidium and molybdenum  are
consistent with in-situ production by an SAGB star.  

We note that, in general, a simple test for ongoing $s$-process activity is to
look for technetium in the atmosphere of the star.  However technetium is also
expected to be produced by the $irp$-process.  The models of \citet{cannon1993}
show Tc number abundances between $10^{-9}$ and $10^{-6}$, which are comparable to
or higher than some of the more Tc-enhanced S-star abundances
\citep[e.g.][]{Abia1998}.  It is in general rather difficult to obtain accurate
abundances for Tc because its spectral lines are blended with those from other
elements.

\section{Lithium}

Lithium is produced during hot bottom burning in the early phases of
AGB and SAGB evolution by the \citet{cameron1955} mechanism.  The typical
temperatures at the base of the convective envelopes of SAGB stars after second
dredge-up are above 60\,MK which allows hot-bottom burning and the synthesis of
lithium.  First $^{7}\rm{Be}$ is produced by the reaction
\begin{equation}
\rm ^3He + {^4He} \rightarrow {^7Be} + \gamma
\end{equation}
and this captures an electron to form $^{7}\rm{Li}$,
\begin{equation}
\rm ^7Be + e^- \rightarrow{^7Li} + \nu.
\end{equation}
Some lithium produced by hot-bottom burning may be transported to
the surface by convection so there is a short period of time in which
Li is significantly enhanced, in many models by more than a factor of ten, at
the surface.  Lithium that is mixed back to or remains in the hot burning region
is subsequently destroyed by proton capture
\begin{equation}
\rm ^7Li + {^1H} \rightarrow 2{^4He}.
\end{equation}
Furthermore, the production of lithium ceases when $^3$He is
depleted, so the surface abundance of lithium decreases after its
initial peak during the early AGB or SAGB phase.

Because lithium is significantly enhanced before any thermal pulses,
the surface composition of an SAGB star is rich in lithium before it
is enhanced in $s$-process isotopes.  In fact, we expect stars with
observed strong Li enhancements not yet to be enhanced in $s$-process
isotopes, for example R~Nor with $A({\rm Li}) = 4.6$
\citep{Uttenthaler2011}.  However, while we do not expect to observe
SAGB or AGB stars with simultaneously strong Li and $s$-process
enhancement, the surfaces of early thermally pulsing SAGB or AGB stars
can have mild Li and $s$-process enhancement at the same time.  During
the first few pulses, light $s$-process isotopes are dredged up to the
surface while most lithium previously produced begins to be destroyed.
Depending on the initial mass and treatment of convection, this period
of mild Li and $s$-process enhancement lasts about $10^4-10^5\,$yr
\citep[see fig.~3 of][]{doherty2014}.  \citet{VanRaai2012} find that
the Li-rich phase lasts for $10^{5}\,$yr in a $6\,\rm M_{\odot}$
model.  Their fig.~7 shows that Li and Rb can be enhanced at the same
time.

Thorne-\.Zytkow objects also synthesise lithium by the same mechanism, the hot
region in this case is the base of the convective envelope just above the
degenerate core.  \citet{Podsiadlowski1995} find that, for a typical example
model, the lithium abundance increases for the first $10^5\,$yr and is still
enhanced after $10^6\,$yr, the maximum lifetime they consider possible.
Therefore there is no problem with simultaneously obtaining rubidium,
molybdenum and lithium in a T\.ZO.

\section{Calcium}

\citet{levesque2014} find enhanced calcium in HV2112.  Calcium is produced by
the capture of alpha particles during the final stages of a star's
nuclear lifetime when photodisintegration and alpha captures convert silicon-28
to nickel-56 immediately before a supernova explosion.  It is not enhanced by
ongoing nucleosynthesis in either T\.ZOs or SAGB stars.  At no point in its
earlier life has an SAGB star reached the necessary conditions to synthesise
calcium anywhere in its interior and so cannot account for this excess calcium
which could then only be explained by fortuitous external pollution.  A T\.ZO
however has more potential because its neutron star core was likely produced by
the collapse of a degenerate iron-rich core that exceeded the Chandrasekhar
mass.  

The total mass of calcium in the Sun, nearly all $^{40}$Ca, is about $6\times
10^{-5}\,\rm M_\odot$.  The metallicity of the SMC is about a tenth that of the
Sun so if we imagine we need to enhance calcium in the an envelope of say
$10\,\rm M_\odot$ or so by a factor of three we require about $10^{-4}\,\rm
M_\odot$ to mix through the whole envelope.

Typical type-II supernovae expel between about $10^{-3}$ and~$2\times
10^{-2}\,\rm M_\odot$ of $^{40}$Ca \citep{woosley1995} which is up to one hundred times what is
needed to pollute our typical T\.ZO envelope.  Unfortunately ejecta of
supernovae leave at over $1{,}000\,\rm km\,s^{-1}$ so that only material
directly impacting the companion can be accreted.  For a separation $a$ and
companion radius $R$ the fraction accreted may be estimated as
\begin{equation} 
f \approx 4\times 10^{-4}\left(\frac{R}{\rm 5\,R_\odot}\right)^2\left(\frac{\rm AU}{a}\right)^2.
\end{equation}
Typical progenitors of type-II supernovae are red supergiants
with radii approaching $1000\,{\rm R_\odot}$ \citep{smartt09}. To fit such a
supergiant into an orbit with our star would require $a > 10\,{\rm AU}$
which means that the fraction of material accreted would be too small even if our
star is already a giant at that stage.

Another possibility is that the calcium could have been synthesised during the
formation of the T\.ZO.  When the degenerate core merges with the neutron star
the conditions in the accretion disc are appropriate for advanced burning and
calcium production.  \citet{metzger2012} has computed the nucleosynthesis
expected during the accretion of the disc produced by a disrupted white dwarf.
He finds that for a $0.6\,\rm M_\odot$ white dwarf, which is the most
similar of his models to the cores of the massive giant companions that we
are considering, around $10^{-3}\,\rm M_\odot$ of calcium should escape from
the disc.  This exceeds the mass required to pollute the envelope to the
extent necessary to produce the surface enhancement of Ca observed in
HV2112.  However material in the disc either accretes directly on to the
neutron star or is expelled in a high-velocity outflow.  The velocity of
this wind from the accretion disc must be roughly equal to the escape
velocity at a distance equal to the characteristic length scale of the disc.
Because the disc is somewhat smaller than the white dwarf radius this
velocity exceeds $10^4\,\rm km\,s^{-1}$.  Naively comparing the kinetic
energy in this outflow with the binding energy of the envelope shows that
there is enough energy available to completely unbind the star, leaving a
naked neutron star.  However, evidence from other objects with relativistic
outflows, including gamma-ray bursts, active galactic nuclei and X-ray
binaries, he suggests that collimated outflows (jets) are common, if not
ubiquitous, in such systems.  \citet{duffell2013} find that for a GRB jet,
somewhat comparable to what we expect for a merging WD--NS system,
Kelvin--Helmholtz instabilities along the boundaries of these jets can mix
as much as a tenth of the material with the envelope they are punching
through.

Hence we postulate that the wind from the central accretion flow was
rather collimated.  In the process of blowing a chimney through the
star convective motions around the boundaries of the flow mixed some
fraction of the wind -- at least a tenth of its mass -- in to the
envelope of the supergiant.  The rest of the wind was lost along the
polar axis of the disc, taking the kinetic energy with it.  This
picture is roughly consistent with what we see in other systems that
posses a relativistic accretion disc and accounts for the calcium
enrichment.

We note also that in Metzger's models only a small fraction of the white
dwarf -- typically about one tenth -- is accreted on to the neutron star.
This is sufficiently little that it is possible for a stable neutron star to
continue as the central object rather than necessitating a collapse to a
black hole.  Because his method does not permit him to follow burning
that occurs whilst the disc is forming he does not consider discs formed by
the disruption of more massive white dwarfs by neutron stars.  Such models
might be more appropriate for the cores of the relatively massive companion
that we postulate here.  Since they would be more massive and denser they
would most likely produce a greater quantity of calcium, reducing the
fraction of the synthesised material which we require the envelope to
absorb.

\section{Conclusions}

We have analysed the possibilities that the unusual SMC supergiant,
HV2112, is either a super-AGB star or a Thorne--\.Zytkow object.
Whilst the formation probabilities are only very slightly in favour of
an SAGB star, the uncertainties in the formation rates are such as to
make this argument very weak.  We find that the majority of HV2112's
observed properties are consistent with either possibility.  The
observed luminosity and temperature are both within those expected.
Both classes of objects are expected to synthesise lithium, rubidium
and molybdenum in situ.  The most distinguishing feature is the high
observed calcium abundance.  This calcium cannot be produced by the
$s$-process, nor can it be accreted directly on to the surface of
HV2112 by the supernova explosion of a binary companion.  We propose
that the calcium could be produced by nuclear burning in the
degenerate core of the companion giant as it merges with a neutron
star to form the Thorne-\.Zytkow object.  The observation of enhanced
calcium thus suggests that HV2112 is most likely a genuine
Thorne-\.Zytkow object and that it has probably formed from the merging of a
binary star progenitor.

\section*{Acknowledgements}
The authors would like to thank Carolyn Doherty and Robert Izzard for
useful discussions.  CAT thanks Churchill College Cambridge for his
fellowship.  RPC is supported by the Swedish Research Council (grants
2012-2254 and 2012-5807) and would like to thank the Institute of
Astronomy, Cambridge for their hospitality during his visit.

\label{lastpage}


\begin{thebibliography}{99}

\bibitem[\protect\citeauthoryear{Abia \& Wallerstein}{1998}]{Abia1998} Abia
C. \& Wallerstein G, 1998, MNRAS, 293, 89

\bibitem[\protect\citeauthoryear{Biehle}{1991}]{Biehle1991} Biehle G. T., 1991,
ApJ, 380, 167

\bibitem[\protect\citeauthoryear{Biehle}{1994}]{Biehle1994} Biehle G. T., 1994,
ApJ, 420, 364

\bibitem[\protect\citeauthoryear{Cameron}{1955}]{cameron1955}Cameron A. G. W.,
1955, ApJ, 121, 144

\bibitem[\protect\citeauthoryear{Cannon}{1993}]{cannon1993}Cannon
  R. C., 1993, MNRAS, 263, 817

\bibitem[\protect\citeauthoryear{Doherty et al.}{2014}]{doherty2014}Doherty 
C. L., Gil-Pons P., Lau H. H. B., Lattanzio J. C., Siess L., 2014, MNRAS, 437, 195

\bibitem[\protect\citeauthoryear{Doherty et al.}{2010}]{doherty2010}
Doherty C. L., Siess L., Lattanzio J. C., Gil-Pons, P., 2010, MNRAS, 401, 1453


\bibitem[\protect\citeauthoryear{Duffell \& MacFadyen}{2013}]{duffell2013}Duffell, P. C., MacFadyen, A. I., 2013, ApJ, 775, 87

\bibitem[\protect\citeauthoryear{Eldridge \&
    Tout}{2004}]{eldridge2004}Eldridge J. J, Tout C. A., 2004, MNRAS,
  348, 201

\bibitem[\protect\citeauthoryear{Garc\'\i a-Berro \& Iben}{1994}]{GarciaBerro1994}
Garc\'\i a-Berro E., Iben I., 1994, ApJ, 434, 306

\bibitem[\protect\citeauthoryear{Glatt et al.}{2010}]{glatt2010}
Glatt K., Grebel E. K., Koch A, 2010, A\&A, 517, 50

\bibitem[\protect\citeauthoryear{Hayashi \& H\={o}shi}{1961}]{hayashi1961}
Hayashi C., H$\bar{\rm{o}}$shi R., 1961, 13, 442

\bibitem[\protect\citeauthoryear{Hurley et al.}{2002}]{hurley2002}
Hurley J. R., Tout C. A., Pols, O. R., 2002, MNRAS, 329, 897

\bibitem[\protect\citeauthoryear{Karakas \&
    Lattanzio}{2014}]{karakas2014}Karakas A. I., Lattanzio J. L.,
  PASA, in press

\bibitem[\protect\citeauthoryear{Kroupa, Tout \& Gilmore}{1993}]{ktg93}
Kroupa P., Tout C. A., Gilmore G., 1993, MNRAS, 262, 545

\bibitem[\protect\citeauthoryear{Lau et al.}{2012}]{lau2012}Lau H. H. B.,
Dougherty C. L., Gil-Pons P., Lattanzio J. C., 2012,
Mem. Soc. Astron. Italy Suppl., 22, 247

\bibitem[\protect\citeauthoryear{Lau et al.}{2011}]{lau2011}Lau H. H. B.,
Dougherty C. L., Gil-Pons P., Lattanzio J. C., 2011, in 
Kerschbaum F., Lebzelter T., Wing R. F., eds, ASP
Conf. Ser. Vol. 445, Why Galaxies care about AGB Stars II:
Shining Examples and Common Inhabitants. Astron. Soc. Pac. San
Francisco, p~45

\bibitem[\protect\citeauthoryear{Levesque et
al.}{2014}]{levesque2014}Levesque E. M., Massey P., \.Zytkow A. N.,
  Morrell N., 2014, MNRAS, in print

\bibitem[\protect\citeauthoryear{Metzger}{2012}]{metzger2012}Metzger B. D., 2012,
MNRAS, 419, 827

\bibitem[\protect\citeauthoryear{Podsiadlowski et al.}{1995}]{Podsiadlowski1995}
Podsiadlowski P., Cannon R. C., Rees M. J., 1995, MNRAS, 274, 485

\bibitem[\protect\citeauthoryear{Smartt et al.}{2002}]{smartt2002}Smartt S. J.,
  Gilmore G. F., Tout C. A., Hodgkin S. T., 2002, ApJ, 565, 1089

\bibitem[\protect\citeauthoryear{Smartt}{2009}]{smartt09}Smartt S. J., 2009,
ARA\&A, 47, 63

\bibitem[\protect\citeauthoryear{van Raai et al.}{2012}]{VanRaai2012}
van Raai M. A., Lugaro M., Karakas A. I., Garc\'\i a-Hern\'andez D. A., and Yong D.,
2012, A\&A, 540, A44

\bibitem[\protect\citeauthoryear{Thorne \&
    \.Zytkow}{1977}]{thorne1977}Thorne K. S., \.Zytkow A. N., 1977,
  ApJ, 212, 832

\bibitem[\protect\citeauthoryear{Thorne \&
    \.Zytkow}{1975}]{thorne1975}Thorne K. S., \.Zytkow A. N., 1975,
  ApJ, 199, 19L


\bibitem[\protect\citeauthoryear{Uttenthaler et al.}{2011}]{Uttenthaler2011}
Uttenthaler S. et al., 2011, A\&A, 531, 88

\bibitem[\protect\citeauthoryear{Vassiliadis \& Wood}{1993}]{vassiliadis1993}
Vassiliadis E., Wood P. R., 1993, ApJ, 413, 641

\bibitem[\protect\citeauthoryear{Woosley \&
    Weaver}{1995}]{woosley1995}Woosley S. E., Weaver T. A., 1995,
  ApJS, 101, 181

\end{thebibliography}
\end{document}